\documentclass[10pt,a4paper,final]{iopart}

\usepackage{graphicx}
\usepackage[breaklinks=true,colorlinks=true,linkcolor=blue,urlcolor=blue,citecolor=blue]{hyperref}
\usepackage{caption}
\begin{document}

\title[Human-brain ferritin studied by muon Spin Rotation: a pilot study]{Human-brain ferritin studied by muon Spin Rotation: a pilot study}

\author{Lucia Bossoni$^{1}$, Laure Grand Moursel$^{2,3}$, Marjolein Bulk$^{2,3,4}$, Brecht G. Simon$^{1}$, Andrew Webb$^{3}$, Louise van der Weerd$^{2,3}$, Martina Huber$^{1}$, Pietro Carretta$^{5}$, Alessandro Lascialfari$^{6}$, Tjerk H. Oosterkamp$^1$}

\address{$^1$Huygens-Kamerlingh Onnes Laboratory, Leiden University, 2333 CA Leiden, The Netherlands}
\address{$^2$Department of Human Genetics, Leiden University Medical Center, Leiden, The Netherlands}
\address{$^3$Department of Radiology, Leiden University Medical Center, Leiden, The Netherlands}
\address{$^4$Percuros BV, Leiden, The Netherlands}
\address{$^5$Department of Physics, Pavia University, Pavia, Italy}
\address{$^6$Dipartimento di Fisica, Universit\`{a} Degli Studi di Milano, Milano, Italy}

\eads{\mailto{bossoni@physics.leidenuniv.nl}}

\begin{abstract}
Muon Spin Rotation is employed to investigate the spin dynamics of ferritin proteins
isolated from the brain of an Alzheimer's disease (AD) patient and of a healthy control, using a sample of horse-spleen ferritin as a reference. A model based on the N\'eel theory of superparamagnetism is developed in order to interpret the
spin relaxation rate of the muons stopped by the core of the protein. Using this model, our preliminary observations
show that ferritins from the healthy control are filled with a mineral compatible with ferrihydrite, while ferritins from the AD
patient contain a crystalline phase with a larger magnetocrystalline anisotropy, possibly compatible with magnetite or maghemite.

\end{abstract}


\vspace{2pc}


\section{Introduction}

Ferritin is a protein that attracts much interest, not only because of its crucial role in iron storage and ferroxidase activity ~\cite{Ward2014,Yang1999}, but also because of its magnetic properties. Ferritin is a nanoscopic hollow protein made of a shell (apoferritin) of molecular weight 450 kDa, containing a core of trivalent iron (Fe(III)) in the mineral form of ferrihydrite, a nano-crystal quite elusive to X-ray diffraction. Ferritin acquires Fe(II), catalyzes iron oxidation, and induces mineralization within its cavity ~\cite{Chasteen1999}. The outer diameter of the shell is 12 nm, regardless of the iron loading, whereas the iron core diameter can vary between 2-3 nm and 7 nm \cite{Gilles2002}, depending on the number of stored ions.\\
\indent It is generally agreed that the ferritin core is antiferromagnetic (AFM) below a temperature in the range of 340-500 K \cite{Gilles2002,Jang2000,Brem2006b}. However, some AFM sublattices do not cancel out completely due to the small particle size. This results in an excess of spin orientation, giving rise to a magnetic moment of 225-400 $\mu_B$ \cite{Brook1998,Gilles2002,Papa2010}. Because of its hexagonal crystal structure (space group $P6_3mc$), the particle possesses a unique easy axis of magnetization \cite{Michel}. The "giant" magnetic moment of the particle can rotate about the crystal easy axis, if it overcomes an energy barrier $E_a$, which depends on the volume of the particle \textit{V} and on the magnetocrystalline anisotropy constant, \textit{K} \cite{Neel}.
If, upon decreasing the temperature, the dynamic time constant of the moment crossing the barrier ($\tau_c$) is greater than the measuring time of the specific experimental technique ($\tau_m$), the magnetic moment is said to be blocked \cite{Fiorani}. This formalism was introduced by N\'eel to describe the magnetism of nanoscopic single-domain particles with an internal magnetic order. The magnetic behaviour of an assembly of these ultra-fine particles was termed superparamagnetism (SPM). This blocking occurs at about 12 K, for ferritin, when $\tau_m \sim 100$ s \cite{Salah}, in a DC magnetometry measurement. In addition to the AFM and SPM phases, it was also proposed that a Curie-Weiss-like behavior could be found at the core-shell interface, as a result of the reduced Weiss field \cite{Brook1998,Neel}. More recently, other models have been proposed, yet the exact spin structure of the nanoparticle and its magnetic properties are still a matter of debate \cite{Brook1998,Silva2009,Gilles2002,Papa2010}. \\
\indent Ferritin has also been extensively studied by neuroscientists, due to its central role in cellular iron homeostasis. Ferritin is at the center of many debates concerning iron toxicity in relation to neurodegeneration, and in particular to Alzheimer's disease (AD). In the brain of AD patients, iron dis-regulation has been reported \cite{Goodman}. This may indicate a malfunction of the storage protein \cite{Dobson2001a}, leading to oxidative stress via Fenton and Haber-Weiss reactions \cite{Ward2014,Toxicol2015,Nunez2012,Crichton}. Microscopy studies revealed the existence of ferritins with a poly-phasic structure: ferritins found in the brain of AD patients seem to contain a higher amount of cubic crystalline phases consistent with magnetite and w\"{u}stite \cite{Quintana2004,Quintana2006}, whereas the 'healthy type'-ferritins may be more abundant in the hexagonal ferrihydride phase. In this scenario, pathological ferritin would be better described by magnetoferritin, an artificial complex made of apoferritin containing a magnetite or maghemite crystal \cite{Moro}. However, these observations were not confirmed by nuclear magnetic resonance (NMR) \cite{Gossuin2005a}. Since magnetoferritin carries a larger magnetic moment than ferrihydrite, i.e. typically between 2000 $\mu_B$ and 9000 $\mu_B$ \cite{Dutta}, one would expect a 200-fold (or larger) increase in the longitudinal and transverse relaxation rates of the water protons surrounding the protein. However, no significant enhancement of the relaxation rate was observed.\\
Additionally, Pan \textit{et al.} showed that in human-liver ferritin, an increasing percentage of octahedrally coordinated Fe(III) migrated to tetrahedral sites and was partially reduced to Fe(II), upon increasing the electron dose in electron microscopy experiments \cite{Pan2006}. Although it is rather unlikely that such alterations would happen only on the 'pathological' ferritin \cite{Quintana2010}, the poly-phasic composition of physiological versus pathological ferritins remains a debated issue. In order to unravel these controversies, here we propose a muon Spin Rotation ($\mu$SR) experiment, as an alternative investigation technique.\\
\indent
$\mu$SR has been successfully employed in the past to study the magnetism of fine-particles systems. In particular, horse-spleen ferritin \cite{Cristofolini,Telling2012} and similar artificial compounds \cite{VanLierop2001} have shown a two-component relaxation of the muon Asymmetry: while the fast-decaying Asymmetry (exponential- or Kubo-Toyabe-like) reflects the static order/collective excitation and the superparamagnetism of the iron core occurring respectively below and above the blocking temperature ($T_B$), the slow exponential tail has been ascribed to the interaction of the muon spin with the protons of the organic shell. However, a study of ferritin purified from human tissue has not been undertaken yet.\\
\indent
In this manuscript, we present a $\mu$SR study of a ferritin sample isolated from the brain of an AD patient, and an age- and gender-matched healthy control (HC). As a third sample and reference, a lyophilized commercial horse-spleen ferritin sample (HoSF) was used.
Firstly, we evaluate the feasibility of $\mu$SR on a human sample. Then we propose a model to interpret the spin dynamics of the ferritin iron core, associated with the SPM effect. Secondly, we draw some preliminary conclusions on the mineral composition of the protein core based on the magnetocrystalline anisotropy constant derived from our model, showing that all steps of the analysis are feasible. Finally, we discuss the limitations and the improvements that should be undertaken in future studies. 

\section{Sample preparation and characterization}

\begin{figure}[ht!]
\centering
\includegraphics[width=8cm,keepaspectratio]{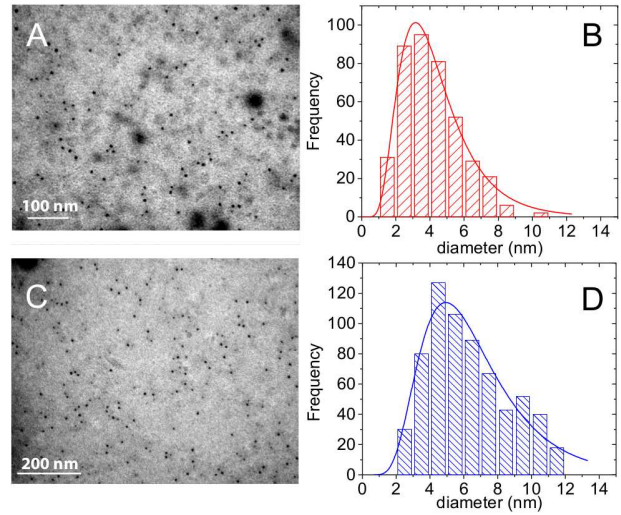}
\captionsetup{justification=justified}
\caption{\textbf{TEM study of the ferritin nanoparticles isolated from human-brain}. (A) TEM micrograph of the ferritin isolated from the Alzheimer's patient. (B) Histogram representing ferritin size distribution from the Alzheimer's patient's ferritin. (C) TEM micrograph of the healthy control's ferritin. (D) Histogram representing ferritin size distribution from the  healthy control's ferritin. The solid line is a fit to the log-normal distribution. Scale bars are 100 nm in panel (A) and 200 nm in panel (C).}
\label{tem}
\end{figure}

\begin{figure}[ht!]
\centering
\includegraphics[width=10cm,keepaspectratio]{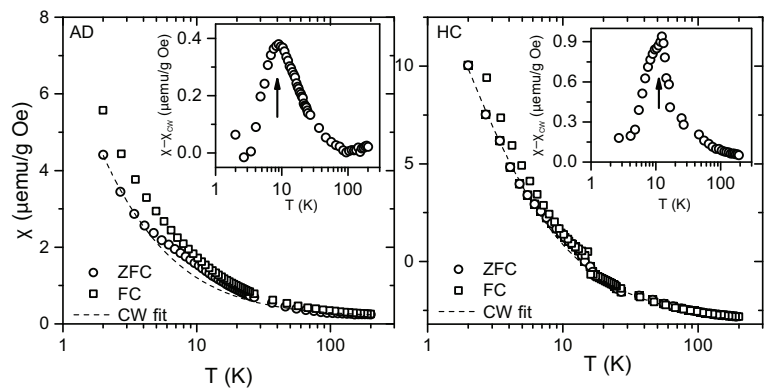}
\captionsetup{justification=justified}
\caption{\textbf{Static Spin Susceptibility of human ferritin, as a function of the temperature.} Zero-Field-Cooled (ZFC) and Field-Cooled (FC) experiments were performed at 50 G. The left panel indicates the data of the Alzheimer's (AD) patient's ferritin, while the right panel refers to the healthy control's (HC) ferritin. The inset shows the ZFC curve, after correction for the Curie-Weiss fit (dashed line), where the peak due to the blocking of the superparamagnetic particles becomes visible. The arrow shows the position of the peak.} 
\label{SQUID_mat}
\end{figure}

One freshly frozen human-brain hemisphere from an AD case (Braak stage 6C, age 90 yrs, female) and one from a healthy age- and gender-matched control (age 88 yrs, female) were obtained from the Netherlands Brain Bank (NBB) of Amsterdam. Patient anonymity was strictly maintained and informed consent was obtained from all prospective donors by NBB in accordance with EU regulations. All tissue samples were handled in a coded fashion, according to Dutch national ethical guidelines (Code for Proper Secondary Use of Human Tissue, Dutch Federation of Medical Scientific Societies).\\
Before the tissue was processed for protein isolation, a section from the temporal region was selected from both the AD and the control individual for an MRI study (see Supplementary Information).
Ferritin used in the $\mu$SR experiment was isolated from 2/3 of the hemispheres of the two individuals by following a modified version of the method reported by Cham \textit{et al}. \cite{Cham1985}. The original protocol requires only a few steps and it is practical when dealing with a large amount of brain material. Moreover, this method retains most of the iron inside the protein, that is our main concern in this experiment.
One part of the brain was homogenized on ice with three parts of Phosphate-buffered saline solution (PBS), with a tissue homogenizer (Omni B style TH Motor 220 V, LA Biosystems). Subsequently, the nuclei and unbroken cells were removed by low-speed centrifugation: 2000 g for 15 min, at 4 $^{\circ}$C. The supernatant was retained and put on ice again. This solution was further diluted with pure methanol 99.8\%  (J. T. Baker, product num.: 8045.2500) in order to reach a final concentration of 40 \% v/v. This new solution was heated for 10 minutes at 75 $^{\circ}$C, and afterward put on ice again, and finally centrifuged at 2000 g for 15 min, at 4 $^{\circ}$C. The ferritin-containing supernatant (final volume: 1.5-2 liters) was retained in order to be further purified, desalted and concentrated with an Amicon Ultra-15 Centrifugal Filter Unit with an Ultracel-100 membrane (UFC910008) with a molecular cutoff of 100 kDa. Aliquots of the supernatants were assayed for ferritin by SDS gel electrophoresis, Western Blot, and Transmission Electron Microscopy.
The remaining solution was freeze-dried for the magnetic characterization. In order to prevent contamination with magnetic material, no metal tools or containers were used in the dissection and handling of the tissue. Ceramic and plastic materials were used instead. \\
\begin{figure}[ht!]
\centering
\includegraphics[width=7cm,keepaspectratio]{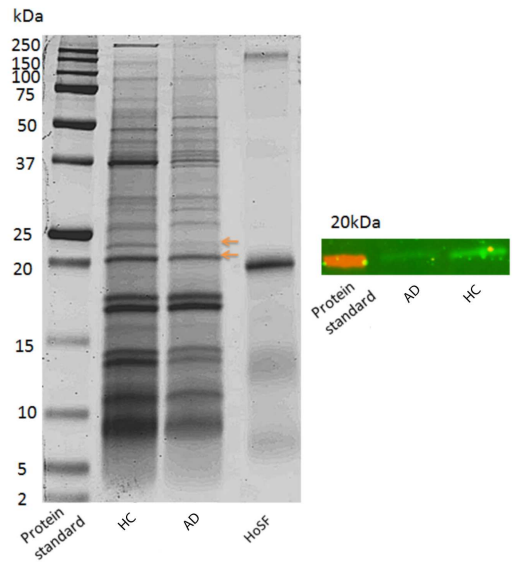}
\captionsetup{justification=justified}
\caption{ \textbf{SDS Page gel and Western Blot of the human ferritin samples together with the horse-spleen ferritin.} (left) SDS Page gel. The heavy ($\sim$ 22 kDa) and light ($\sim$ 19 kDa) chains of human-brain ferritin are shown by the orange arrows. AD refers to the Alzheimer's patient, HC to the healthy control, and HoSF to the horse-spleen ferritin sample. (right) Western Blot of the same human sample showing a bright green band corresponding to the heavy chain of ferritin.}
\label{gel}
\end{figure}
SDS-gel electrophoresis was done in parallel to Western Blot analysis: the results show the presence of both ferritin sub-bands in the human samples (Fig. \ref{gel}).\\
One part of the gel was stained to assess purity and protein concentration, while the other part was used for Western Blot to determine the presence and molecular weight of ferritin. Protein solutions were loaded on the lanes of a 4-20\% Criterion TGX\textsuperscript{TM} pre-cast gel purchased from Bio-Rad. As a reference, Precision Plus Protein\textsuperscript{TM} Dual Xtra Prestained Protein Standards (product num.: 1610377) and ferritin from equine spleen purchased from Sigma-Aldrich (productr num.: F4503) were employed. Horse-spleen ferritin was loaded as a reference sample in three different concentrations. The gel was run using a standard Laemmli sample buffer and Tris/glycine/SDS running buffer. The gel was stained with PageBlue\textsuperscript{TM} protein staining solution (ThermoFisher Scientific, product num. 24620) and imaged with the Li-Cor Odyssey$^{TM}$ imaging system. The human ferritin concentration was estimated from the heavy and light chain bands: 227.3 $\mu$g/g for the AD sample and 210.5 $\mu$g/g for the HD. Taking into account the solution volume after the concentration step, and the powder obtained after freeze-drying, we estimated that the ferritin concentration loaded onto the $\mu$SR sample holder was 11 \% and 14.3 \% of the total sample mass for the AD and HC ferritin, respectively. However, a spread of values in the concentration of different batches cannot be excluded.\\
\indent 
The presence of ferritin was verified by Western Blot. Anti-human-ferritin goat antibody NBP1-06985 purchased from Novus Biologicals was used as a primary antibody. Fluorescent donkey antigoat antibody was used as a secondary antibody. Antibody solutions and blocking solutions were prepared with tris-buffered saline Tween (TBST) and non-fat 5\% powder milk.
The nitrocellulose membrane was imaged with the same Li-Cor Odyssey$^{TM}$ imaging system. \\
\indent 
In order to confirm the presence of the ferritin iron core, we performed Transmission Electron Microscopy (TEM) on the purified protein solution. TEM images were acquired using a JEOL 1010 Transmission Electron Microscope, at 70 kV. A 10 $\mu l$-solution containing human-brain ferritin was drop-cast on a carbon coated copper grid (200 mesh, Van Loenen Instruments) and let dry for a few minutes. The solution that did not absorb into the grid was removed with dust-free paper. The ferritin solution obtained after the isolation protocol was imaged before freeze-drying (Fig. \ref{tem}) and later the freeze-dried powder used for the $\mu SR$ experiment was re-suspended in milliQ water and imaged with the same protocol. No appreciable changes in the iron core size distribution, as a result of freeze-drying and $\mu$SR experiments, were observed (data not shown). Fig. \ref{tem} shows dark dots of electron-dense material indicating the iron-filled ferritin cores. The size distribution of the ferritin particles, obtained by the analysis of several TEM micrographs, follows a log-normal curve. Moreover, our results suggest that the AD ferritins have a smaller core size, with a mean of 4.39 nm and a standard deviation of 2.17 nm, than the HC ferritins which show a mean-core size of 6.64 nm and a standard deviation of 3.11 nm. These values are reasonably in agreement with the literature \cite{Galvez}. No larger, i.e. several tens of nm, particles were found in the analyzed solution by TEM inspection. \\
\indent 
Freeze-dried powders of human-brain ferritin were characterized with a 7 T-MPMS SQUID magnetometer. In these experiments, the static spin susceptibility is measured at 50 G after cooling the sample in two different ways. In the Zero-Field-Cooled (ZFC) case, the sample is cooled from room temperature, down to 2 K without applying a magnetic field. Then the field is increased to 50 G and the static susceptibility is measured. In contrast, in the Field-Cooled (FC) case, the sample is cooled with the field of 50 G applied continuously. The static spin susceptibility ($\chi$) of the human ferritin samples is shown in Fig. \ref{SQUID_mat}. The ZFC $\chi$ shows the typical peak of a superparamagnet after subtracting a Curie-Weiss trend from the data \cite{Brem2006c,Salah} (inset Fig. \ref{SQUID_mat}). The AD sample $\chi$ shows a broad peak, centered around 8.8 $\pm$ 0.9 K, indicating a blocking temperature ($T_B $) smaller than that observed in the HoSF sample (12 $\pm$ 1 K, $\chi$ data are shown elsewhere \cite{Bossoni16}). $T_B $ of the HC sample is the same as the HoSF sample. These values of $T_B $ suggest that the average core size of the AD sample is smaller than the one of the healthy control and horse-spleen samples. This observation is in agreement with the TEM characterization. 
Please note that the FC curve does not show this peak, as 50 G is enough to block the superparamagnet at low temperatures. Furthermore, the presence of a peak in the ZFC curve in the range of $\sim$ 9-12 K rules out the possibility that a substantial part of our sample contains magnetite particles of tens of nm \cite{Kirschvink,Maher,Pankhurst2008,Bossoni16}.\\
Finally, the presence of iron in all the powder samples was confirmed by Laser-ablation Inductively coupled Mass spectrometry.

\section{Results}

Ferritin powders of different origin were loaded onto a kapton holder, with a thin (25 $\mu$m) kapton window to allow the muons to reach the sample. Experiments were carried out in longitudinal geometry, both in Zero-Field and longitudinal field mode. The experiments were performed at the GPS beamline of the Paul Scherrer Institute, where a continuous wave (CW) muon beam is available. CW muons allow a high time resolution, which in turn enables strongly inhomogeneous internal fields to be probed.
A forward and a backward detector were used at all times. The time-dependent Asymmetry A(t) function is calculated as:
\begin{equation}
A(t) = \frac{F(t)-\alpha B(t)}{F(t)+\alpha B(t)}
\end{equation}
where F(t) and B(t) are the positron counts in the forward and backward detector respectively, while $\alpha$ is an instrumental parameter which compensates for the difference in efficiency between the two detectors. Our estimated $\alpha$ ranged between 0.787 and 0.859 and it was calibrated in a transverse field of 50 G, at high temperature (170 K). 
Data were bunched with the 'Constant error' binning option of the Wimda software, in which the bin length exponentially increases with time from the initial value so that the counts per bin and the resulting error remained fixed. Data were fitted between 0 and 6 $\mu s$.
Different fitting functions were tested, such as the Uemura function \cite{Uemura1985}, various combinations of Lorentzian and Gaussian Kubo-Toyabe \cite{KT}. The dynamical Kubo-Toyabe and Keren functions \cite{Keren} were also tested, but they could not capture all the features of the data, as the empirical model discussed in this manuscript. Muon Spin Rotation raw data were processed with Wimda, and fitted with Wimda, Matlab 2016a and OriginLab(R) Pro 2016.

\subsection{Horse-spleen ferritin Zero-Field Asymmetry}

\begin{figure}[ht!]
\centering
\includegraphics[width=11cm,keepaspectratio]{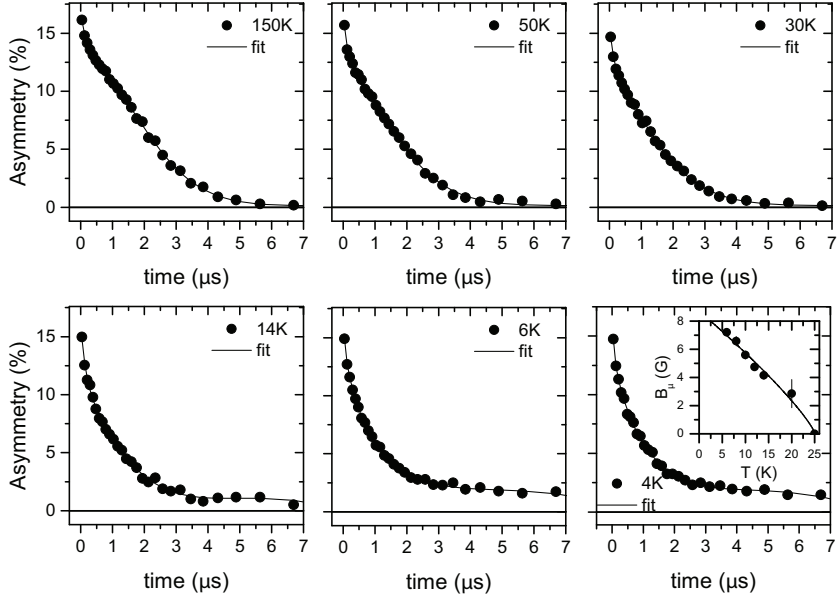}
\captionsetup{justification=justified}
\caption{\textbf{Asymmetry decay of the horse-spleen ferritin at different temperatures.} The dots represent binned raw data acquired in Zero-Field and the solid line is the fit according to Eqs. \ref{eq1} and \ref{eq2}. The recovery of the Asymmetry at longer times is visible in the bottom row of plots. The $A=0$ line is shown for reference. The temperature is shown in each panel. The inset is the fitted internal field $B_{\mu}$ responsible for the increase in the Asymmetry at long times. The line is a guide to the eye. For further details on the fit, please refer to the Supplementary Information.}
\label{Fig_horse}
\end{figure}

Before carrying out the $\mu$SR study on the human-brain ferritin, a sample of freeze-dried horse-spleen ferritin was investigated. Horse-spleen ferritin was chosen because it is similar to human ferritin and therefore a good reference.
Zero-Field $\mu$SR on HoSF shows a change in the muon Asymmetry decay $A(t)$ as a function of the temperature (Fig. \ref{Fig_horse}). 
We observed an initial Asymmetry ($A_0$) of 15 \%, which equals 2/3 of the full Asymmetry. This signal loss will be discussed later in this section.\\
In the Zero-Field measurement data, two regimes were identified. At temperatures higher than 20 K, the Asymmetry is well described by the sum of two contributions, namely an exponential term, decaying in the first 300-500 ns, and a slower stretched-exponential component, evolving from a Gaussian at high temperature, to a single exponential at low temperature. The decay of $A(t)$ was modeled by assuming that a fraction of muons probes the local magnetic field of the protein shell ($f_{shell}$), and the rest probes the core ($f_{core}$):
\begin{equation}
A(t) = A_0 ( f_{core} e^{-\sigma t} + f_{shell} e^{-(\lambda t)^{\beta}}) +B
\label{eq1}
\end{equation}
We observed a fraction ratio of  $\frac{f_{core}}{f_{shell}} \sim \frac{1}{4}$. A similar value was found by Cristofolini \textit{et al.}, who suggested that such a fraction may originate from the core/shell mass ratio \cite{Cristofolini}. If we consider the mean values for the diameter of the core (5-6 nm) and of the shell (12 nm), we would expect to observe $\frac{f_{core}}{f_{shell}}\sim 0.08-0.14$, instead. The discrepancy was ascribed to conformational changes of the protein shell, as suggested for example for freeze-dried proteins \cite{Roy}. Note that the experimental ratio could be different also because the muon implantation is not random. \\
In Eq. \ref{eq1}, $\sigma$ represents the spin-lattice relaxation rate of the 'core muons', analogous to the NMR $1/T_1$ \cite{book1,Lascia}. As it will be shown, the fast decaying component is weakly dependent on the application of a longitudinal magnetic field.
Finally, $\lambda$ is the spin relaxation rate of the muons probing the protein shell, and the stretched exponent in Eq. \ref{eq1} can be associated with either a distribution of muon sites or with an anisotropic hyperfine coupling, leading to a distribution of relaxation rates. \\
Below 20 K, the Asymmetry shows an increase at times longer than $\sim$ 4 $\mu$s (Fig. \ref{Fig_horse}), indicating the onset of a static magnetic phase. We note that in contrast to the work of Cristofolini \textit{et al.} on HoSF \cite{Cristofolini}, here the Gaussian decay is not reached at any temperature below 20 K. The best fit to the data was obtained by adding a slow oscillating term, with a small amplitude ($f_{sm} \sim$ 6\%):
\begin{equation}
A(t) = A_0 ( f_{core} e^{-\sigma t} + f_{shell} e^{-(\lambda t)^{\beta}}+f_{sm}\cos(2\pi \gamma B_{\mu}+\phi)) +B,
\label{eq2}
\end{equation}
where the baseline was deduced from the high-temperature regime, $\gamma = 8.516 \times 10^4$ rad $s^{-1} G^{-1}$ is the muon gyromagnetic ratio, $B_{\mu}$ is the local field probed by the muons, $\phi$ is a phase constant and $f_{sm}$ represents the fraction of 'shell muons' probing static magnetism. The internal field $B_{\mu}$ of $\sim$ 8 G, at low temperature (Fig. \ref{Fig_horse} (inset)) can be explained in terms of the local dipolar field produced by blocked iron magnetic moments of 5 $\mu_B$, at an average distance of 1.8 nm from the muon implantation site (See the Supplementary information for a more detailed description of the fitting). This internal field is of the same order of magnitude as that derived by Cristofolini \textit{et al.} \cite{Cristofolini}, on horse-spleen ferritin, at low temperature.

\begin{figure}[ht!]
\centering
\includegraphics[width=12cm,keepaspectratio]{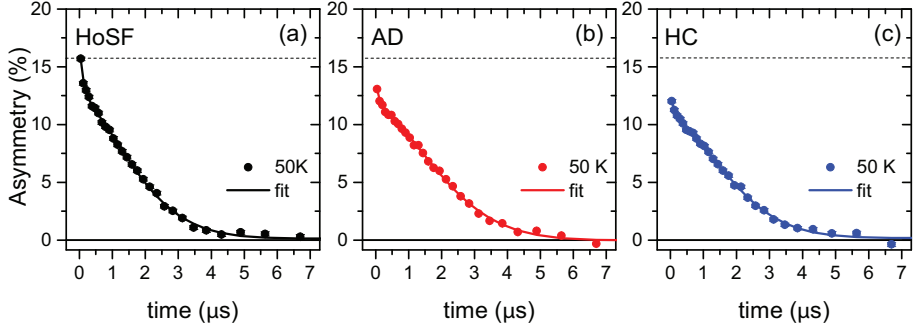}
\captionsetup{justification=justified}
\caption{\textbf{Asymmetry decay for the horse-spleen ferritin and human-brain ferritin at 50 K.} Equation \ref{eq1} fits equally well the three data sets. Human data have been baseline corrected. Baseline = 2.1 for the healthy control (HC) and 3.54 for the Alzheimer's (AD) sample, respectively. The Asymmetry of the horse-spleen ferritin \textbf{(a)} is marked with a dashed line and reported in the two other panels \textbf{(b)} and \textbf{(c)}, for comparison. The dots represent binned raw data and the solid line is the fit according to Eq. \ref{eq1}. The $A=0$ line is shown for reference.}
\label{Fig_human}
\end{figure}

\subsection{Human-brain ferritin Zero-Field Asymmetry}

\begin{figure}[ht!]
\centering
\includegraphics[width=11cm,keepaspectratio]{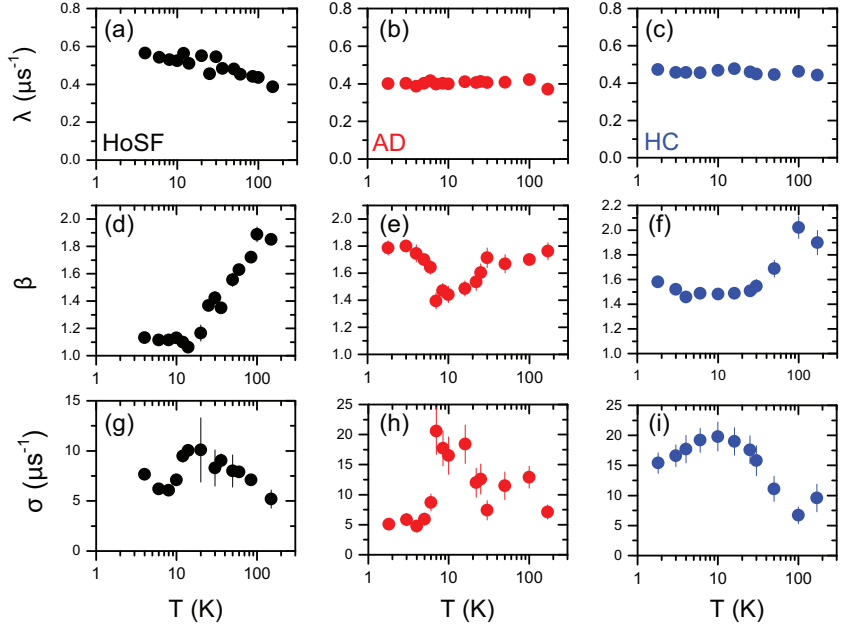}
\captionsetup{justification=justified}
\caption{\textbf{Best fitting parameters for the horse and human ferritin data sets.} Muon spin relaxation rate $\lambda$, stretched exponential $\beta$  and spin-lattice relaxation rate $\sigma$ obtained from the fit of different data sets: for HoSF, see panels \textbf{(a)}, \textbf{(d)} and \textbf{(g)}; for the AD patient's ferritin, see panels \textbf{(b)}, \textbf{(e)} and \textbf{(h)}; for the healthy control's ferritin (HC), see panels \textbf{(c)}, \textbf{(f)} and \textbf{(i)}. The error bars represent the standard error of the estimated coefficients.}
\label{fit_results_MM}
\end{figure}

The Asymmetry decay of the human-brain ferritin is remarkably similar to that of HoSF, as far as the high-temperature limit is concerned (Fig. \ref{Fig_human}), therefore we fitted the data with the high-temperature model of Eq. \ref{eq1}. However, by a more careful inspection of the data, some differences can be observed between the human and the horse data sets. The most remarkable are that the human ferritin Asymmetry does not increase at long times and at low temperature, as described by Eq. \ref{eq2}. This may be due to the lower ferritin concentration in the human sample. Secondly, the stretched exponent does not decrease below 1.4 in both human samples, showing that the mono-exponential limit, in the slow terms, is never reached (Fig. \ref{fit_results_MM}). Finally, the initial Asymmetry of the human data is smaller than in the HoSF data (Fig. \ref{Fig_human}), as it will be discussed in the next section. \\
Fig. \ref{fit_results_MM} summarizes the best-fit parameters for the relaxation rates ($\lambda,\sigma$) and stretched exponent ($\beta$), as a function of the temperature, for the three samples.
The Asymmetry decay of human ferritins was described by a model with three fitting parameters, while the ratio of the core-shell muon fraction was fixed. The baseline was also left free to vary, but it did not show significant variations with the temperature.

\subsection{Longitudinal $\mu$SR field measurements on human-brain and horse-spleen ferritin}

\begin{figure}[ht!]
\centering
\includegraphics[width=11cm,keepaspectratio]{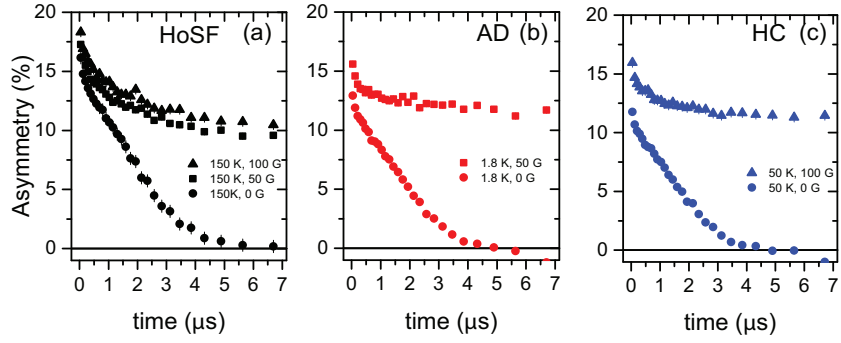}
\captionsetup{justification=justified}
\caption{\textbf{Asymmetry decay for different applied longitudinal fields and different temperatures of horse and human ferritin.} \textbf{(a)} Asymmetry measured at Zero-Field and increasing field on the horse-spleen ferritin (HoSF) sample, at high temperature, showing the quenching of the stretched exponential term. \textbf{(b)} Asymmetry measured at Zero-Field and 50 G on the Alzheimer's (AD) human ferritin sample, at 1.8 K, showing that both the initial Asymmetry and the slow-decaying term are affected by the increase of the longitudinal field. \textbf{(c)} Asymmetry measured at Zero-Field and 100 G on the human healthy control (HC) ferritin sample, showing a similar behavior, at 50 K. Data have been baseline corrected. }
\label{LF}
\end{figure}

\indent
Longitudinal-Field (LF) $\mu$SR on the horse and human ferritin show two effects in the muon Asymmetry. Firstly, the application of a weak field enhances the initial Asymmetry $A_0$, as also observed by other authors in a HoSF sample \cite{Cristofolini,Telling2012}. Secondly, the stretched exponential relaxation is partially quenched by the longitudinal field (Fig. \ref{LF}). 
The loss of $A_0$ was earlier ascribed to muonium formation (a radioactive isotope of hydrogen) after the $\mu^+$ captures an electron, as often occurs in organic matter and in polymers \cite{BlundellEPL,Pratt}. In this case, the polarization can be fully recovered by the application of a longitudinal magnetic field, large enough to decouple the hyperfine interaction between the electron and the muon.
Given the large amount of organic material in our sample and the partial recovery of $A_0$ with the application of an LF, it is reasonable to assume that the signal loss at $t=0$ is due to muonum formation. Moreover, since the human sample is about seven times more dilute than the HoSF, one may ascribe the larger loss of initial Asymmetry to a larger fraction of other proteins in the human sample (Fig. \ref{Fig_human}).\\
The quenching of the stretched exponential relaxation by small fields is in agreement with the hypothesis that this decay originates from the muons implanting in the protein shell, and probing weak dipolar fields of the proton nuclei. Moreover, the fast exponential term is not affected by the LF, thus suggesting that in this case, the local field is either static and strong (i.e. a few Tesla) or dynamic, as expected for the iron core.

\section{Discussion}

\begin{figure}[ht!]
\centering
\includegraphics[width=12cm,keepaspectratio]{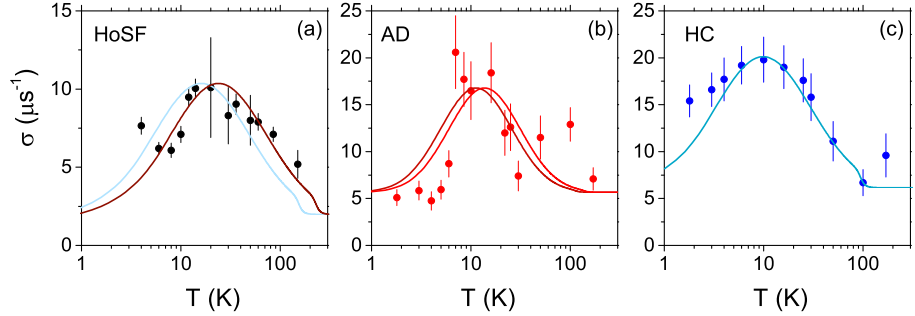}
\captionsetup{justification=justified}
\caption{\textbf{Comparison between simulation and experimental data of the spin-lattice relaxation rate $\sigma$.} \textbf{(a)} Experimental $\sigma$ for the horse-spleen ferritin and simulated $\sigma$ according to the model discussed in the text and $K$ = 2.5 K$/$nm$^3$ (light blue curve) and 2.7 K$/$nm$^3$ (dark brown curve). \textbf{(b)} Experimental $\sigma$ for AD ferritin and simulated curve for $K = 4.7$ and $5.7$ K$/$nm$^3$ (red solid curves). Here the AD particle size distribution is taken into account. \textbf{(c)} Experimental $\sigma$ for HC ferritin and simulated curve for $K=2$ K$/$nm$^3$ (blue curve).}
\label{Sim_Exp}
\end{figure}

Among the different magnetic resonance techniques, muon Spin Rotation benefits from an almost 100 \% spin-polarized muon beam, leading to a signal which is not limited by the Boltzmann polarization. 
Since the spin probe (the muon itself) is implanted into the sample, the technique is neither limited by the sensitivity nor by the natural abundance of the internal probe, as it is in the case of NMR. Additionally, the signal intensity is not dependent on the strength of the external magnetic field. In $\mu$SR, Zero-Field experiments are possible and the muon current is so small ($\sim$ pA) that no or minor sample damage is expected. \\
Moreover, muon Spin Rotation offers several advantages for the study of nanoscopic magnetic systems: it directly probes the stray field produced by the metallic nanoparticle, and it captures its spin dynamics in a broad dynamic range (between 10$^{-12}$ s and 10$^{-5}$ s). On the other hand, obtaining structural information from relaxation rates still remains a challenge.
In the next section, we propose a model to interpret the spin-lattice relaxation rate of the muon fraction stopping in the iron core of the protein ($\sigma$).\\ 
\indent
There are at least two channels for the spin relaxation of muons implanting in ferritin: (i) spin excitations of the AFM ordered lattice, which would appear at low temperatures, in the form of a power law \cite{Pincus,Bossoni2011} and (ii) N\'eel relaxation of the particle giant moment, activated by thermal energy. The correlation time for the latter spin fluctuations is typically described by the N\'eel-Arrhenius relation, for non-interacting particles \cite{Fiorani}:
\begin{equation}
\tau_c(T,V)= \tau_0 e^{E_a/k_B T}
\label{eq3}
\end{equation}
where $1/\tau_0$ is the attempt frequency, $E_a = V K$ \cite{Note1,Silva2008} is the energy barrier, and $k_B$ is the Boltzmann constant. Such jumps of the particle magnetic moment are random between opposite values, therefore the spin-spin correlation function becomes \cite{Slichter}:
\begin{equation}
<S_z(0)S_z(t)>= S_0^2 e^{-t/\tau_c(T,V)}
\label{eq4}
\end{equation}
Hence, one can borrow the formalism from the NMR theory of the spin-lattice relaxation rate, for a mono-dispersed particle distribution \cite{book1}:
\begin{eqnarray}
\sigma(T,V) &\propto & \gamma^2<h_0^2>\int{<S_z(0)S_z(t)>e^{-i \omega_L t}dt} \nonumber   \\
 &=& \gamma^2<h_0^2>\frac{\tau_c}{\tau_c^2\omega_L^2+1} 
\label{eq5}
\end{eqnarray}

where $<h_0^2>$ is the static \textit{rms} value of the hyperfine field, $\omega_L  = \gamma B_{\mu}$ and $B_{\mu}$ is the local field probed by the muons implanting in the core \cite{Arosio}. Since the ferritins' diameters are log-normally distributed, as observed by TEM, Eq. \ref{eq5} should be integrated over the volume distribution:
\begin{equation}
\sigma(T) \propto \int_{V_0}^{V_1} \sigma(T,V) \frac{\exp(-\frac{(\ln(V)-<\ln(V)>)^2}{2 \Delta^2})}{\Delta V \sqrt{2 \pi}} dV
\label{eq6}
\end{equation}
where $\Delta$ and $<\ln(V)>$ are respectively the standard deviation and the mean of the log-normal distribution.
\begin{figure}[ht!]
\centering
\includegraphics[width=10cm,keepaspectratio]{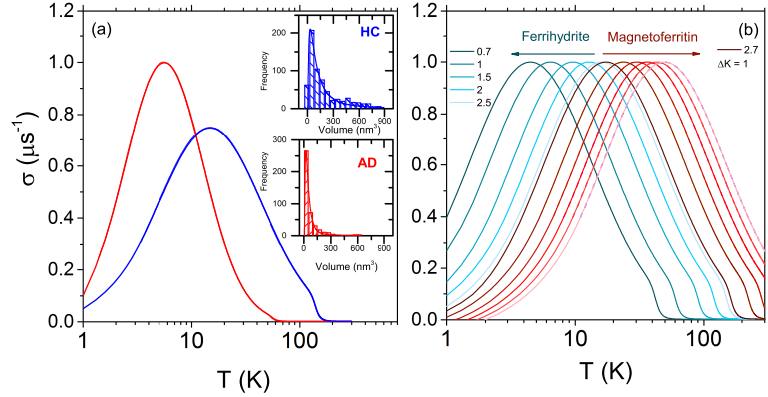}
\captionsetup{justification=justified}
\caption{\textbf{Simulation of the spin-lattice relaxation rate $\sigma$.} \textbf{(a)} Simulated spin-lattice relaxation rate, using the volume particle distributions taken from electron microscopy (TEM) data (inset). The simulated $\sigma$ for the AD sample is shown in red and the HC in blue. Parameters used for the simulation are $K$ = 2.3 K/nm$^3$ (ferrihydrite), $\tau_0 =$ 0.1 ps \cite{Moro,Madsen2008},  $B_{\mu}= 8$ G. The insets report the experimental distributions of the core-particle size from TEM and fit according to the log-normal distribution (solid curve). \textbf{(b)} $\sigma$ simulation for different $K$ values compatible with ferrihydrite (solid lines in shades of blue) and magnetoferritin (in shades of red). Here only one particle volume distribution, compatible with the HC, is considered. $K$ values used for the simulations are reported in units of K/nm$^3$. For the magnetoferritin simulations, the range of $K$ values is 2.7-7.7 K/nm$^3$, in steps of 1 K/nm$^3$. The amplitude of the peaks is normalized.
}
\label{Sim_vol}
\end{figure}
In order to interpret $\sigma$, we choose $\tau_0$ and $K$ from literature values for ferrihydrite (see Fig. \ref{Sim_vol} and its caption).\\
The simulation in Fig. \ref{Sim_vol} \textbf{(a)} shows that particles with a larger volume (as observed for HC) display a peak in $\sigma$ at  higher temperature, and that particles with a broader particle-size distribution display a broader peak in $\sigma$. Fig. \ref{Sim_vol} \textbf{(b)} shows a positive correlation between the value of $K$ and the temperature of the peak in $\sigma$. Here, instead of choosing one single value for $K$, we present a range of values, as reported in literature. Indeed, $K$ depends on several factors such as chemical composition, kind of magnetic ion, particle shape, amount of frustrated spins at the particle surface, distribution of nucleation sites inside the core, and level of crystallinity. Plausible values for $K$, in the case of the ferrihydrite mineral, range from 0.7 to 2.5 K/nm$^3$ \cite{Nolte,Gunther,Li2009}, while for magnetoferritin, and magnetite nanoparticles they can be found between 2.7 and 9.7 K/nm$^3$ or higher \cite{Martinez,Moro}. However, intermediate values of 2.97 K/nm$^3$ have also been measured for HoSF \cite{Madsen2008}. In Fig. \ref{Sim_vol} \textbf{(b)} the chosen $K$ values are limited to 7.7 K/nm$^3$, for illustration purposes. Moreover, in all the following simulations, $B_{\mu}$ has been fixed to 8 G. This assumption, based on the experimental observation on the HoSF data, may be an over-simplification in the case of the human data.\\
A comparison between the experimental and simulated $\sigma$ is shown in Fig. \ref{Sim_Exp}, for the three investigated samples. 
All the samples show a broad peak in $\sigma$, in the same temperature range observed in other works on HoSF and its synthetic analogues \cite{Cristofolini,VanLierop2001}. In the case of HoSF, the acceptable values for $K$ fall between 2.5 and 2.7 K$/$nm$^3$ (Fig. \ref{Sim_Exp} \textbf{(a)}), which is in line with the literature findings for HoSF, as discussed above. We note that $K$ may be overestimated, if the actual mean particle size was larger than considered here. 
The HC sample shows a broader peak centered in a temperature region compatible with $K\sim 2$ K$/$nm$^3$, which suggests the presence of a single ferrihydrite phase (Fig. \ref{Sim_Exp} \textbf{(c)}), in agreement with the expectations for ferrihydrite and possibly characterized by a lower degree of crystallinity with respect to the HoSF. Finally, the AD sample shows a narrower peak in $\sigma$, which results from a narrower particle-size distribution (Fig. \ref{Sim_Exp} \textbf{(b)}), and a significantly larger anisotropy constant of $K \sim 4.7-5.7$ K$/$nm$^3$, corresponding to a $\sim$ 63 \% increase of the magnetic anisotropy, which is more than what can be expected from a difference in particles size between AD and HC \cite{Martinez}. This constant falls in the experimentally observed range of magnetoanisotropy for magnetoferritin nanoparticles. \\
\indent
It is worth stressing that this is the first attempt to use muon Spin Rotation to determine the mineralization form of the ferritin-iron core in human-brain ferritin, as a complementary tool to electron microscopy. Therefore, in order to improve the current data, and confirm our preliminary results, the human ferritin concentration and purity should be improved. A possibility would be to add a further purification step, such as affinity chromatography, at the end of our protocol. As many of these steps can be further improved, future experiments are certainly feasible and will be sensitive enough to determine differences with greater accuracy.

\section{Conclusions}

We isolated ferritin from the brain of an AD patient and a healthy age- and gender-matched individual. The protein solution was characterized by biochemical and physical techniques assessing the protein concentration and verifying the superparamagnetic properties, and the presence of iron in the sample. The characterization shows that the isolation protocol needs further improvements, in order to significantly increase the concentration and purity of the ferritin solution. \\
$\mu$SR was then used to probe the spin dynamics of the iron core of human-brain ferritin, as a function of the temperature. Our pilot experiment showed Asymmetry spectra that significantly resembled those of commercial and highly pure horse-spleen ferritin, which was used as a reference in this study. A broad peak in the spin-lattice relaxation rate of the muons stopping in the core allowed us to identify the blocking temperature of ferritin. We proposed a model to interpret the spin-lattice relaxation rate, based on the N\'eel theory of superparamagnetism, and on the experimental size distribution of the obtained ferritin particles. The comparison between simulation and experimental data gave us an indication of the magnetocrystalline anisotropy constant of the ferritin iron core. Our analysis suggests that ferritin isolated from the control subject is in agreement with a mineral phase compatible with 'physiological' ferrihydrite, while the ferritin isolated from the Alzheimer's patient contains an iron mineralization form with a larger $K$ constant, in the range observed for magnetoferritin, when the size distribution obtained from electron microscopy is taken into account. However, to draw further conclusions about the composition of ferritin in relation with AD, ferritin isolated from different AD patients should be investigated. This pilot study shows that $\mu$SR is a promising approach to the study of iron-loaded human-brain proteins.

\section*{Acknowledgements}
We are grateful to M. de Wit, A. Amato and C. Baines for proving help during the muon Spin Rotation experiment and for useful discussion. We thank L. Cristofolini and J. Wagenaar for fruitful discussions. G. Lamers, J. Willemse, L. van der Graaf, J. Aarts, N. Lebedev and C. Koeleman for technical assistance during ferritin freeze-drying and characterization. We thank W. Breimer for helping during the MRI data acquisition.\\
This work was supported by the Dutch Foundation for Fundamental Research on Matter (FOM), by the Netherlands Organization for Scientific Research (NWO) through a VICI fellowship to T. H. O. and through the Nanofront Program. One of us (M. B.) was supported by the FP7 European Union Marie Curie IAPP Program, BRAINPATH, under grant number 612360.
Partial funding was provided by European Research Council, Advanced Grant 670629 NOMA MRI.

\section*{References}
\bibliographystyle{iopart-num}

\bibliography{biblio_musr}

\end{document}